\documentclass[aps,showpacs,twocolumn,superscriptaddress]{revtex4}

\usepackage{graphicx}
\usepackage{dcolumn}
\usepackage{amsmath}

\newcommand{\vp}{{\vec p}}

\begin{document}

\preprint{\parbox[t]{45mm}{\small ANL-PHY-9964-TH-2001\\ UNITU-THEP-01/19}}

\title{Pair Creation and an X-ray Free Electron Laser}

\author{R.~Alkofer}
\affiliation{Institut f\"ur Theoretische Physik, Universit\"at T\"ubingen, Auf
der Morgenstelle 14, D-72076 T\"ubingen, Germany\vspace*{1ex}}
\author{M.B.~Hecht}
\affiliation{Physics Division, Bldg 203, Argonne National Laboratory, Argonne
Illinois 60439-4843\vspace*{1ex}}
\author{C.D.~Roberts}
%\email[]{cdroberts@anl.gov}
%\homepage[]{http://www.phy.anl.gov/theory/staff/cdr.html}
\affiliation{Physics Division, Bldg 203, Argonne National Laboratory, Argonne
Illinois 60439-4843\vspace*{1ex}}
\author{S.M.~Schmidt}
\affiliation{Institut f\"ur Theoretische Physik, Universit\"at T\"ubingen, Auf
der Morgenstelle 14, D-72076 T\"ubingen, Germany\vspace*{1ex}}
\author{D.V.~Vinnik}
\affiliation{Institut f\"ur Theoretische Physik, Universit\"at T\"ubingen, Auf
der Morgenstelle 14, D-72076 T\"ubingen, Germany\vspace*{1ex}}

%\date{\today}

\begin{abstract}
\rule{0ex}{3ex}
Using a quantum kinetic equation coupled to Maxwell's equation we study the
possibility that focused beams at proposed X-ray free electron laser
facilities can generate electric field strengths large enough to cause
spontaneous electron-positron pair production from the QED vacuum.  Our
approach yields the time and momentum dependence of the single particle
distribution function.  Under conditions reckoned achievable at planned
facilities, repeated cycles of particle creation and annihilation take place
in tune with the laser frequency.  However, the peak particle number density
is insensitive to this frequency and one can anticipate the production of a
few hundred particle pairs per laser period.  Field-current feedback and
quantum statistical effects are small and can be neglected in this
application of non-equilibrium quantum mean field theory.
\end{abstract}

\maketitle

%%%%%%%%%%%%%%%%%%%%%%%%%%%%%%%%%%%%%%%%%%%%%%%%%%%%%%%%%%%%%%%%%%%%%
%\section{Introduction}
%%%%%%%%%%%%%%%%%%%%%%%%%%%%%%%%%%%%%%%%%%%%%%%%%%%%%%%%%%%%%%%%%%%%%%
%\label{one}

The QED vacuum is unstable in the presence of a strong external field and
decays by emitting electron-positron pairs.  The pair production rate was
first calculated for a static, homogeneous electric field in the early part
of the last century~\cite{Sauter} and since then many aspects have been
studied in detail.  Using the Schwinger formula~\cite{Sauter} one finds that
a sizeable rate requires a field: $E_{cr}:= m_e^2/e = 1.3 \times
10^{18}\,$V/m, which is very difficult to produce in the laboratory.  (We use
$\hbar = c= 1$.)

The possibility of spontaneous pair creation from the vacuum is of particular
interest in ultra-relativistic heavy ion collisions~\cite{signals,bastirev}.
Since the QCD string tension is large ($\sqrt\sigma \sim 400\,$MeV), flux
tube models of the collision generate a background field that is easily
strong enough to initiate the production process via a Schwinger-like
mechanism.  Feedback between the external field and the field created by the
produced particles' motion drives plasma oscillations.  This is the
back-reaction phenomenon, which has been much
discussed~\cite{Back,bloch,vinnik}.  While this Vlasov-equation-based
approach has met with some phenomenological success, a rigorous justification
in QCD is wanting.

Pair creation by laser beams in QED has also been
discussed~\cite{Brezin,Popov}, and proposed X-ray free electron laser (XFEL)
facilities at SLAC~\cite{slac} and DESY~\cite{desy}, which could generate
field strengths~\cite{Chen,Ringwald} $E \approx 0.1 E_{cr}$, promise to
provide a means to explore this phenomenon.

Vacuum decay is a far-from-equilibrium, time-de\-pen\-dent process and hence
kinetic theory provides an appropriate descriptive framework.  For spatially
homogeneous fields, a rigorous connection between kinetic theory and a
mean-field treatment of QED has been established~\cite{basti,kme}.  The
derivation makes plain that the true kinetic equation's source term is
intrinsically non-Markovian, and this is expressed in properties of the
solution~\cite{basti,kme,rau,sms}.  Herein we use this quantum Vlasov
equation to obtain a description of the time evolution of the momentum
distribution function for particles produced via vacuum decay at the planned
XFEL facilities.

A gauge and Lorentz invariant description of an electromagnetic (e.m.) field
is obtained using
\begin{equation}
{\cal F} = \frac 1 4 F_{\mu\nu}F^{\mu\nu} = \frac 1 2 (\vec E^2 - \vec B^2),
\quad
{\cal G} = \frac 1 4 F_{\mu\nu}\widetilde F^{\mu\nu} = \vec E \cdot \vec B.
\end{equation}
An e.m.\ plane wave always fulfills ${\cal F} = {\cal G} =0$ and such a
light-like field cannot produce pairs~\cite{Troup}.  Therefore, to produce
pairs it is necessary to focus at least two coherent laser beams and form a
standing wave.  Subsequently we assume an ``ideal experiment:'' owing to the
diffraction limit the spot radius of the crossing beams cannot be smaller
than the wavelength, so we choose $r_\sigma \approx \lambda$; and we assume a
space volume in which the electric field is nonzero but the magnetic field
vanishes.  Even for carefully chosen X-ray optical elements, such a situation
is impossible to achieve in practice, which means that the field strength
actually available to produce particles is weaker than the peak field value
and hence our estimate of the production rate will be an upper bound.

Our idealised experiment is realised through a vector potential $A_\mu(t)$
that generates an antiparallel electric field $\vec E_{ext}(t) =
-d\vec{A}(t)/dt$: this is the background field, which in temporal gauge,
$A_0=0$, reads $\vec{A}(t) = (0,0,A(t))$, and the alternating laser field is
then
\begin{equation}\label{laserE}
{\vec E}_{ext}(t) = (0,0,E_0 \sin(\Omega t));\,\,\,\, {\vec B} = 0,
\end{equation}
where $\Omega=2\pi /\lambda$ is the frequency.  (Table~\ref{table} provides
our laser field parameters.)  The total electric field is $E(t) = E_{ext}(t)
+ E_{int}(t)$ where the internal field is induced by the back-reaction
mechanism.  As we will show, $E_{int}$ can be neglected under anticipated
XFEL conditions.

\begin{table}[t]
\caption{\label{table} Laser field parameters are specified in columns one
and two: Set~I is XFEL-like; Set~II is strong.  Columns three and four
describe the density, $n_{max}(t_>)$, and total number of produced particles,
$N(t_>) = \lambda^3 n_{max}(t_>)$, Eq.~(\protect\ref{Nt1}), where $t_>$ is
the time at which the number density reaches its (for weak fields, local)
maximum.  A typical laser pulse length is $\sim 80\,$fs~\protect\cite{desy}.}
\vspace*{1ex}
\begin{ruledtabular}
\begin{tabular*}
{\hsize} {l@{\extracolsep{0ptplus1fil}}
|l@{\extracolsep{0ptplus1fil}}l@{\extracolsep{0ptplus1fil}}
|l@{\extracolsep{0ptplus1fil}}r}
%{l|c|c|c}
          & $\lambda\,$(nm) & $E_0\,$(V/m) & $n_{max}(t_>)\,$(fm$^{-3})$
          &$N(t_>)$ \\\hline    
%         & & & & \\
Set ~Ia    & 0.15     & $1.3\times 10^{17}$ 
&$4.6\times 10^{-13}$ & $\sim 10^3$ \\
Set ~Ib    & 0.075     & $1.3\times 10^{17}$ 
&$4.6\times 10^{-13}$ &$\sim 10^2$\\
Set IIa   &  0.15    & $1.3\times 10^{18}$ 
&$7.2\times 10^{-10}$ & $\sim 10^{6}$  \\
Set IIb   &  0.075    & $1.3\times 10^{18}$ 
&$6.4\times 10^{-10}$ &$\sim 10^{5}$\\
\end{tabular*}
\end{ruledtabular}
\end{table}

A key quantity in the description of nonequilibrium particle production
processes is the single-particle momentum distribution function, $f(\vp,t)$.
For Dirac particles coupled to an Abelian gauge field it can be obtained by
solving a quantum Vlasov equation~\cite{basti,kme}.  Once $f(\vp,t)$ is known
the calculation of the produced particle number density and total particle
number is straightforward.  In general the Vlasov equation involves source
and collision terms, and its coupling to Maxwell's equation provides for the
field-current feedback typical of plasmas.  The absolute and relative
importance of these terms depends on the magnitude of the background field
and the mass of the produced particles~\cite{bastirev}.  For the relatively
weak XFEL-like fields we expect the produced-particle number density to be
small and hence collisions to be rare.  Therefore we neglect the collision
term and arrive at the following pair of coupled equations:
\begin{eqnarray}
\label{KE}
\frac{d\, f(\vp,t)}{d\,t} &=&  
\frac{eE(t)\varepsilon_\perp^2}{2\omega^2 (\vp,t)}
\int_{t_0}^t dt'\frac{eE(t')\big[1-2f(\vp,t')\big]}{\omega^2 (\vp,t')}
\nonumber \\
&& \qquad \qquad \times 
\cos\left[ 2\int_{t'}^t d\tau ~\omega (\vp,\tau) \right] ,
\\
{\dot E}_{int}(t) &=& -2 e
\int\!\frac{d^3p}{(2\pi)^3}\,\frac{p_\|}{\omega(\vp)}
\Biggl( f(\vp,t) +
\nonumber \\
&&\frac{\omega^2(\vp)}{eE(t)p_\|}\frac{d\, f(\vp,t)}{d\,t} 
 - \,\frac{e \dot E(t)\,\varepsilon_\perp ^2}{8\,
\omega^4(\vp)p_\|} \Biggr) ,
\label{ME}
\end{eqnarray}
with the three-vector momentum $\vp=(\vp_{\perp} ,p_\parallel)$; the
transverse mass-squared $\varepsilon_{\perp}^2=m_e^2+p_{\perp}^2$; and the
total energy-squared $\omega^2(\vp,t)=\varepsilon_{\perp}^2+p_\parallel^2$;
$p_{\parallel}(t)=p_3 - e\,A(t)$.

The effect of quantum statistics on the particle production rate is evident
in the factor ``$[1 - 2 \,f]$'' in Eq.~(\ref{KE}), which ensures that no
momentum state has more than one spin-up and one spin-down fermion.  In
addition, both this statistical factor and the ``$\cos$'' term introduce
non-Markovian character to the system: the first couples in the time history
of the distribution function's evolution; the second, that of the field.  One
anticipates that such memory effects are only important for very strong
fields: $E\sim E_{cr}$; i.e., when the time-scale characteristic of the
produced particles' Compton wavelength is similar in magnitude to the time
taken to tunnel through the barrier~\cite{kme,rau,sms}.  We will verify that
such effects are unimportant for XFEL-like fields and that, in this
application, one may employ the low density (l.d., $f\ll 1$) approximation to
distribution function:
\begin{eqnarray}
f^{\rm l.d.}(\vp,t) &=& \int_{t_0}^t d\tau \frac{eE(\tau)
\varepsilon_\perp^2}{2\omega^2 (\vp,\tau)} \int_{t_0}^\tau
dt'\frac{eE(t')}{\omega^2 (\vp,t')} \nonumber \\ 
&& \qquad \qquad \times \cos\left[ 2\int_{t'}^t d\tau' ~\omega (\vp,\tau')
\right].
\label{ld}
\end{eqnarray}

Equations~(\ref{KE}) and (\ref{ME}) are coupled: Eq.~(\ref{KE}) depends on
the total field, $E(t)$, calculated from Eq.~(\ref{ME}), and the electric
field itself depends on the polarisation and conduction currents, which are
proportional to $df/dt$ and $f$, respectively.  For strong fields the
internal currents are non-negligible and field-current feedback will generate
plasma oscillations.  However, as we will demonstrate, that is not relevant
for realisable XFEL fields.

We turn now to a quantitative analysis of two exemplary electric field
strengths in Eq.~(\ref{laserE}), solving Eqs.~(\ref{KE}) and (\ref{ME}) using
a $4^{th}$ order Runge-Kutta method.  The weaker field: $E=0.1\,E_{cr}$;
i.e., Sets I in Table~\ref{table}, should be obtainable at the proposed XFEL
facilities~\cite{slac,desy}.  Sets II in the table, obtained with $E=E_{cr}$,
provide a strong field comparison.
%  and furthermore, given the rapid development in laser technology, may be
%  realisable in the not too distant future.

\begin{figure}[t]
\centerline{\includegraphics[height=6.5cm,angle=-90]{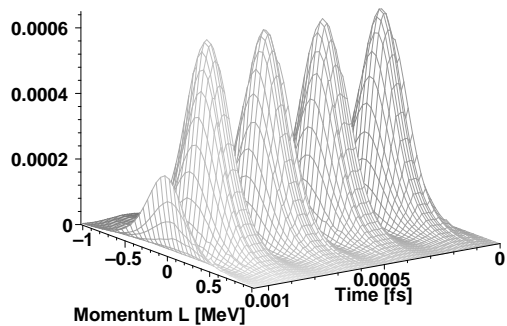}}
\caption{\label{fig1} Time and momentum dependence of the distribution
function $f(\vp,t)$ for $p_\perp=0$ obtained with the conditions described in
Set Ia of Table~\protect\ref{table}.  This alternating laser field produces
repeated cycles of particle creation and annihilation.}
\end{figure}

Figure~\ref{fig1} depicts the time evolution of the momentum distribution
function for electrons produced under currently anticipated XFEL conditions.
This relatively weak alternating field generates repeated cycles of pair
creation, with about $1.6\times 10^5$ local maxima in one laser pulse.  The
production rate: $S(\vp,t) = df(\vp,t)/dt$, is clearly time-dependent and
vanishes at these local maxima.  Therefore an estimation of $S$ at a fixed
time, as in Refs.~\cite{Chen,Ringwald}, is potentially misleading.  Under
these conditions there is no accumulation of produced particles because
creation and annihilation processes are balanced; i.e., all the particles
produced have sufficient time to annihilate again before the next cycle of
particle creation.  Hence the number of particles cannot exceed the value at
the first of the repeated maxima.

\begin{figure}[t]
\centerline{\includegraphics[height=6.5cm,angle=-90]{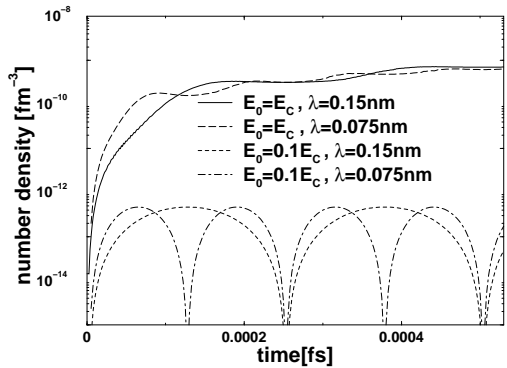}}
%\centerline{\epsfig{figure=df_comparN.eps,height=6.5cm,angle=-90}}
\caption{\label{fig2} Time evolution of the number density for Sets I and II,
Table~\protect\ref{table}.  In strong fields, particles accumulate, leading
to the almost complete occupation of available momentum states.  In weak
fields, repeated cycles of particle creation and annihilation occur in tune
with the laser frequency: $\Omega_{\rm Ia}=2.0 \times
10^{18}\,2\pi s^{-1}$; $\Omega_{\rm Ib}=4.0 \times
10^{18}\,2\pi s^{-1}$.}
\end{figure}

The situation is markedly different for strong fields, as made plain by
Fig.~\ref{fig2}, which depicts the time evolution of the number density:
\begin{equation} 
n(t)=\int \frac{d^3p}{(2\pi)^3}f(\vp,t)\,.
\end{equation}  
When $E=E_{cr}$ the interval between bursts of particle creation is so small
that not all the particles can annihilate before the next cycle of particle
production begins.  Therefore the particles accumulate until almost all
momentum states are completely occupied.  In this case field-current feedback
effects may become important.

Having calculated the time-dependent number density we can estimate the
number of particles produced in the spot volume via vacuum decay.  Assuming
the diffraction limit on focusing can be reached then $r_\sigma = \lambda$,
$V_\sigma \approx \lambda^3$, and the peak particle number is
\begin{equation}
\label{Nt1}
N(t_>) \approx \lambda^3\, n(t_>)\,,
\end{equation}
where $t_>$ is the time-location of the peak number density (which is any of
the local maxima for XFEL fields).  Our results are presented in
Table~\ref{table}.  We remark that, following a very different route,
Ref.~\cite{Chen} also arrives at a value similar to our Set Ia estimate of
$\sim 10^3$ particles.  Increasing the peak field strength by one order of
magnitude increases the produced particle yield by $10^3$, as is obvious from
Fig.~\ref{fig2}.

\begin{figure}[t]
\centerline{\includegraphics[height=6.5cm,angle=-90]{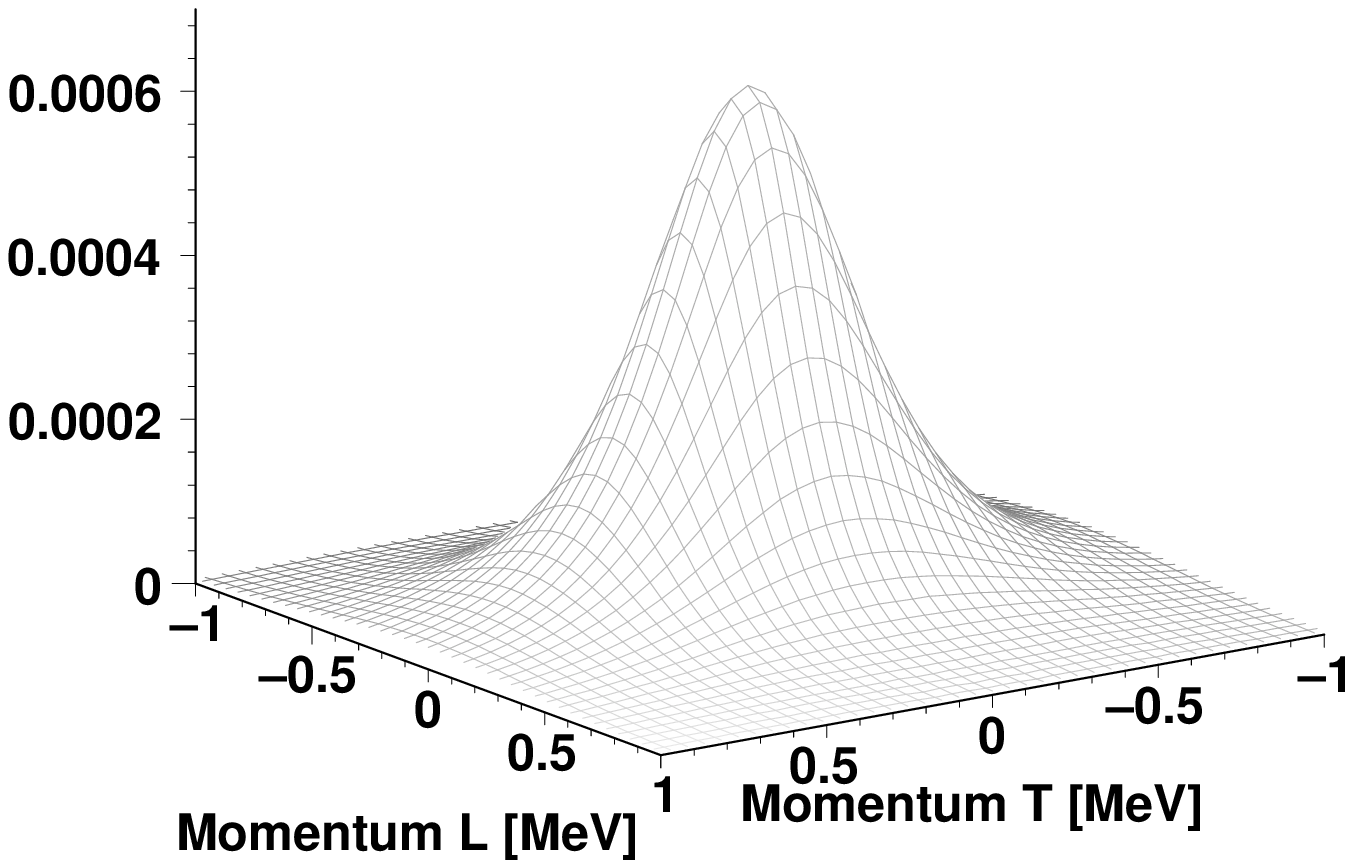}}
\vspace*{1ex}

\centerline{\includegraphics[height=6.5cm,angle=-90]{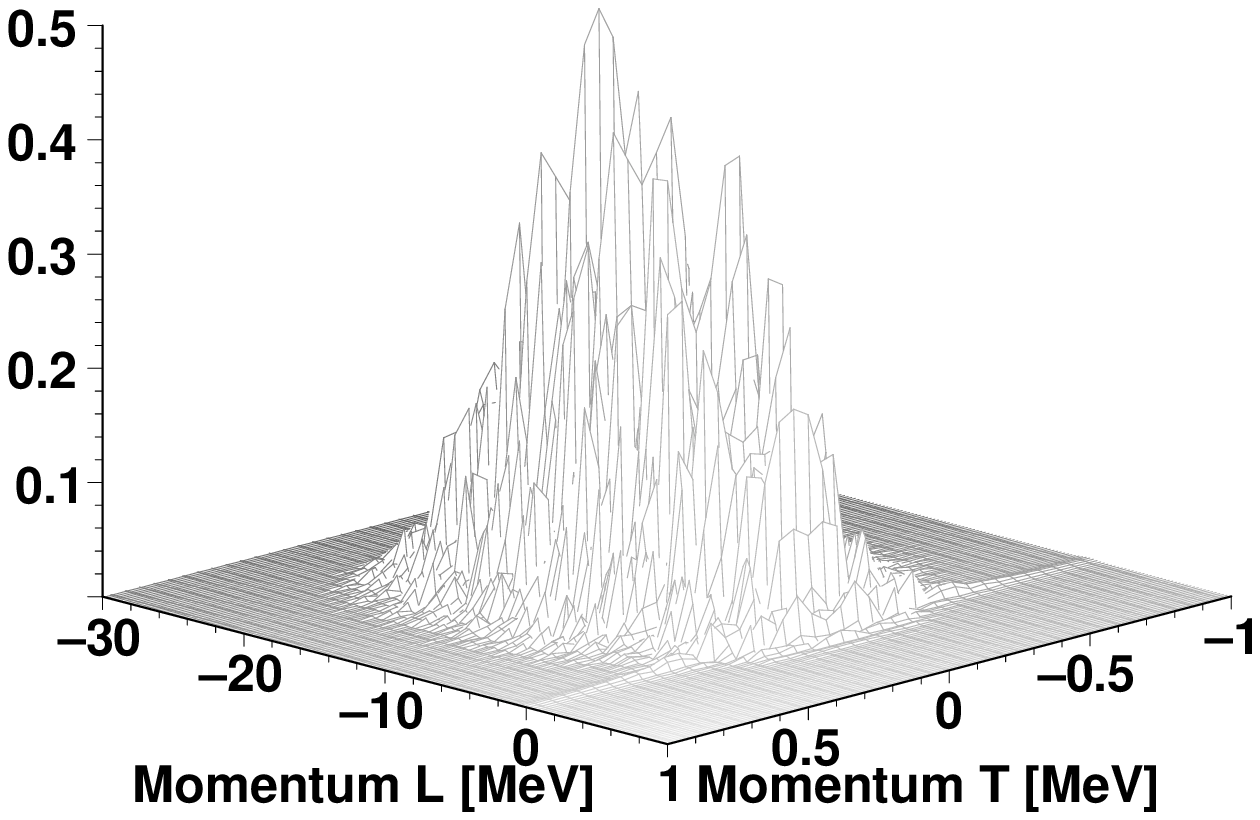}}
%\centerline{\epsfig{figure=pp_01_b30.eps,height=4.5cm,angle=-90}\epsfig{figure=pp_1_b30n.eps,height=4.5cm,angle=-90}}
\caption{\label{fig3} Momentum-dependence of the distribution function: Upper
panel, Set Ia; Lower panel Set IIa, Table~\protect\ref{table}.  Note the
vastly different magnitudes.  The accessible range of longitudinal momenta is
magnified for strong fields and the irregular structure expresses the
influence of non-Markovian effects in the source term for this case.}
\end{figure}

Our complete numerical solution also provides the full momentum distribution,
which is depicted in Figs.~\ref{fig3}.  For both weak and strong fields the
maximum transverse momentum is approximately $1\,$MeV but the maximum
longitudinal momentum is much larger for strong fields.  This result owes
itself to the minimal coupling of the vector potential to the parallel
momentum component but is invisible for weak fields.  Furthermore the
presence of non-Markovian effects for strong fields (see discussion of
Eq.~(\ref{KE})) is evident in the irregular structure in the lower panel cf.\
the smooth distribution in the XFEL-like field distribution depicted in the
upper panel.

\begin{figure}[t]
\centerline{\includegraphics[height=6.5cm]{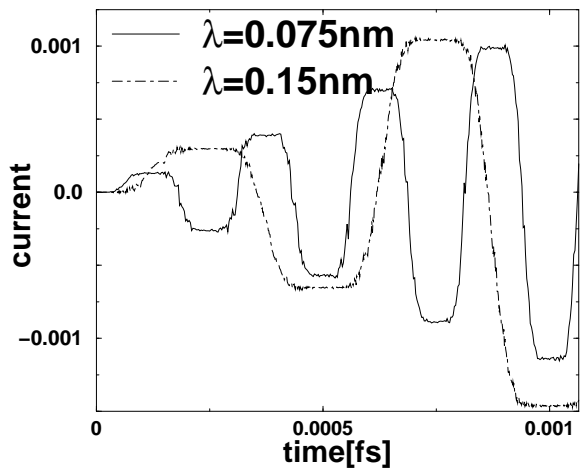}}
%\centerline{\epsfig{figure=current.eps,height=5.0cm,angle=-90}}
\caption{\label{fig4} Current for Sets II of Table~\protect\ref{table}.  The
plateaux occur when the produced particles reach their terminal velocity,
almost the speed of light.  The oscillatory behaviour is driven by the laser
frequency. (NB.\ For an XFEL-like field the current is zero on the scale of
this figure.)}
\end{figure}

In Fig.~\ref{fig4} we plot the time evolution of the current.  For an
XFEL-like field the induced current is zero on the scale of this figure.
(NB.\ It is a factor of $10^3$ smaller than the external current.)  The
plasma oscillation period in this system can be estimated using the
ultrarelativistic formula
\begin{equation}
\tau_{\rm pl}^{UR} \approx \frac{\sqrt{n_{max} m_e + E_{cr}^2}}{n_{max} e/2}
\approx 6.5\times 10^{-19}\,s
\end{equation} 
cf.\ the laser periods: $2.5\times 10^{-19}\,$s and $5.0\times 10^{-19}\,$s,
which shows that the oscillatory behaviour evident in the figure is
prescribed by the laser frequency.  (NB.\ It emphasises, however, that were
the amplitude of the induced current larger, field-current feedback may
become important.)  The behaviour of the strong field current is instructive.
The focused laser beams generate a volume of high field strength in which
particles are created.  The strong field persists long enough for all the
particles to be accelerated to their maximum velocity, almost the speed of
light in this case, which explains the appearance of the plateaux: the
current saturates when all the produced particles reach their terminal
velocity.  The plateaux end as the alternating laser field changes sign and
accelerates the electrons in the opposite direction.  Each successive plateau
is higher because, as explained in connection with Fig.~\ref{fig2}, for
strong fields the period between particle production cycles is small and
hence there is a net accumulation of particles.

We have explored the possibility of electron-positron pair production through
vacuum decay at the focus of achievable X-ray free electron laser beams via
the self-consistent solution of a quantum Vlasov equation coupled to
Maxwell's equation.  This is a rigorous application of non-equilibrium
quantum mean field theory.  Our complete calculation confirms that the effect
will be observable if proposed facilities achieve their design goals.  Under
such conditions the production rate is time-dependent, with repeated cycles
of particle production and annihilation in tune with the laser frequency, but
the peak particle number is independent of the laser frequency: up to $10^3$
pairs may be produced in the spot volume.  Our procedure yields the full
single particle momentum distribution function (Fig.~\ref{fig3}) and hence
provides comprehensive phase space information about the $e^+ e^-$ pairs.  It
also verifies that, at present design goals: collisions among the produced
particles can be neglected because the maximum particle number density is low
(Fig.~\ref{fig2}); a Schwinger-like source term is adequate because the
non-Markovian features of the true source term have no impact
(Fig.~\ref{fig3}) and hence the low density approximation to the distribution
function, Eq.~(\ref{ld}), is accurate; and field-current feedback will not
influence the production process (Fig.~\ref{fig4}).

\begin{acknowledgments}
We thank A.\ Ringwald and W. Dittrich for helpful discussions.  This work was
supported by the US Department of Energy, Nuclear Physics Division, under
contract no.~\mbox{W-31-109-ENG-38}; the Deutsche
For\-schungs\-ge\-mein\-schaft, under contract no.\ SCHM~1342/3-1; and
benefited from the resources of the US National Energy Research Scientific
Computing Center.
\end{acknowledgments}

%%%%%%%%%%%%%%%%%%%%%%%%%%%%%%%%%%%%%%%%%%%%%%%%%%%%%%%%%%%%%%%%%%%%%%

\end{document}